\documentclass[
amsmath,amssymb,
prx,
nobalancelastpage,
raggedbottom,
twocolumn
]{revtex4-2}

\usepackage{graphicx}
\usepackage{float}
\usepackage{dcolumn}
\usepackage{bm}
\usepackage{hyperref}
\usepackage[mathlines]{lineno}
\usepackage{siunitx}
\usepackage{braket}

\begin{document}

\title{Trapped fermions frustrate multiterminal Josephson networks}

\author{Antonio Manesco}
\email{trilattices@antoniomanesco.org}
\author{Karsten Flensberg}
\affiliation{Niels Bohr Institute, University of Copenhagen, DK-2100 Copenhagen, Denmark
}

\date{\today}

\begin{abstract}
Coherent control of fermions has become a standard approach for engineering quantum devices.
Here, we explore parity-induced frustration in multiterminal Josephson arrays.
We focus on odd-parity Josephson trijunctions, where the presence of a trapped fermion induces a phase winding in each trijunction.
Connecting these trijunctions constrains their shared phases, resulting in a long-distance phase coherence where the frustrated configurations dominate the low-energy physics. 
In the classical regime, these arrays map onto the Kagome XY antiferromagnet, whose established Berezinskii-Kosterlitz-Thouless (BKT) transition yields algebraically decaying $6e$ phase correlations.
We show that the $6e$ phase coherence is manifested as a pronounced third harmonic in the current-phase relation of these arrays.
While quantum phase fluctuations suppress the $6e$ coherence, we show that through spin-orbit coupling they also lead to spin correlations and entanglement between the trapped fermions.
Our results establish fermion-induced frustration as a route to composite superconducting coherence and phase-mediated spin entanglement in superconductor-semiconductor artificial quantum materials.
\end{abstract}

\maketitle

\section{\label{sec:level1} Introduction}

Superconductor-semiconductor hybrids combine the macroscopic quantum coherence of superconductors with the electric tunability of semiconductors.
This combination allows investigating how the control of microscopic fermionic degrees of freedom affects the collective superconducting behavior.
Coherent manipulation of fermionic states motivated the development of applications such as Majorana qubits~\cite{kitaev_unpaired_2001,leijnse_parity_2012,zatelli_majorana_2026, tsintzis_majorana_2024, alicea_non-abelian_2011}, Andreev spin qubits~\cite{hays_coherent_2021,padurariu_superconducting_2010,pita-vidal_direct_2023,hays_continuous_2020,pita-vidal_strong_2024}, and fermionic quantum computation~\cite{vilkelis_fermionic_2024}.
These platforms also enabled studies of engineered topological superconductivity~\cite{oreg_helical_2010,lutchyn_majorana_2010,leijnse_parity_2012, noauthor_realizing_nodate, dvir_realization_2023}, coherent Cooper pair splitting~\cite{wang_singlet_2022, recher_andreev_2001, hofstetter_cooper_2009, herrmann_carbon_2010}, and chiral Andreev transport~\cite{day_chiral_2025,dahl_directional_2026,day_probing_2026}.
More recently, advances in this field have allowed the engineering of hybrid superconductor/semiconductor superlattices, leading to the observation of superconductor-insulator, Berezinskii-Kosterlitz-Thouless (BKT), and quantum phase transitions in these systems~\cite{bottcher_berezinskii-kosterlitz-thouless_2024, bottcher_superconducting_2018,mikalsen_observation_2026,bondar_two_2026,wang_long-range_2026}.
These developments motivate a research question connecting the two scales: how can the control of microscopic fermionic states be used to engineer collective behavior in a superconducting network?

\begin{figure}[!ht]
    \centering
    \includegraphics[width=\linewidth]{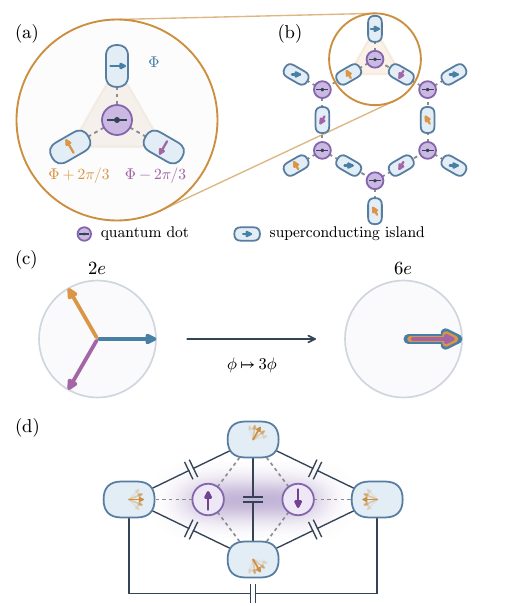}
    \caption{
    (a) Frustrated trijunction: an odd-parity quantum dot (purple) couples three superconducting islands (blue).
    The filled-level marker denotes one occupied quasiparticle; arrows indicate condensate phases separated by $2\pi/3$.
    (b) Honeycomb array of trijunctions with shared superconducting islands.
    The highlighted trijunction is enlarged in (a).
    (c) Composite $6e$ coherence: tripling the three phases of $\Psi_{2e}\propto e^{i\phi}$ aligns them in the composite field $\Psi_{6e}\propto\Psi_{2e}^{3}\propto e^{i3\phi}$.
    (d) Phase-mediated spin entanglement of two spin doublets.
    }
    \label{fig:intro}
\end{figure}

In this work, we explore a route to frustration in Josephson networks driven by changes in the parity of a quantum dot coupled to conventional superconductors~\cite{feinberg_spontaneous_2014}.
Frustration in superconducting networks often relies on explicit time-reversal symmetry breaking~\cite{teitel_josephson-junction_1983, delfino_cooper-pair_2025} or combining superconducting materials with different pairing symmetries~\cite{Smilde2002,Hilgenkamp2003}.
We illustrate this idea with the three-terminal Josephson junction shown schematically in Fig.~\ref{fig:intro}(a), where the odd-parity state favors a $\pi$ phase difference between each pair of superconducting leads~\cite{van_dam_supercurrent_2006, sharov_influence_2005}.
The resulting frustration of the Josephson potential leads to two degenerate energy minima, in which the superconducting phases wind by $\kappa 2\pi/3$ across successive pairs of leads around the junction, forming a discrete vortex with chirality $\kappa=\pm 1$~\cite{feinberg_spontaneous_2014, manesco_loopless_2026}.
While this Josephson frustration has been previously proposed as a route for flux control of the energy-phase relation in loopless qubit architectures~\cite{manesco_loopless_2026}, here we explore how correlations emerge when we connect these trijunctions in a network.

In this work, we explore networks of these frustrated multiterminal Josephson junctions, such as the honeycomb lattice shown in Fig.~\ref{fig:intro}(b).
Connecting these trijunctions constrains the superconducting phases of the shared islands, resulting in a highly degenerate manifold of compatible phase configurations.
Averaging over these multiple configurations suppresses the usual $2e$ phase coherence, while the phase coherence of the composite field $\Psi_{6e}\propto\Psi_{2e}^3$ is preserved, as illustrated in Fig.~\ref{fig:intro}(c).
In the classical regime, the emerging $6e$ phase coherence in two-dimensional honeycomb arrays is a result of a BKT transition, similar to previous predictions in Kagome materials and arrays~\cite{park_superconducting_2001,Song2025}.
We show that the composite phase coherence manifests in the current-phase relation of finite arrays as a pronounced third harmonic, which can be probed in transport measurements.

Quantum phase fluctuations suppress the $6e$ coherence, but open a route to entangle the fermionic states trapped in the quantum dots.
Because spin-orbit couples the spin of the trapped fermion to the phase of its neighboring islands, spin polarization and trijunction chirality are locked~\cite{manesco_loopless_2026}.
We show that this locking leads to a long-range antiferromagnetic coupling between the trapped fermions in two trijunctions sharing superconducting islands.
As we allow quantum phase fluctuations, the ground state of the two-trijunction device becomes a highly entangled spin state.
These results establish frustrated multiterminal Josephson networks as a platform for composite superconducting coherence and spin entanglement between distant quantum dots.

\section{Parity-induced phase frustration}
\label{sec:single-trijunction}

\begin{figure}[!t]
    \centering
    \includegraphics[width=\linewidth]{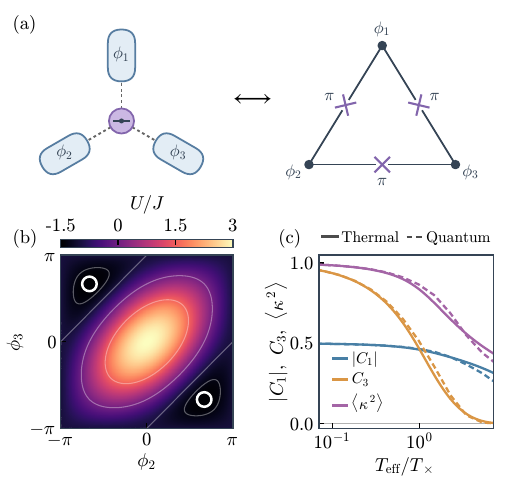}
    \caption{
        (a) A frustrated Josephson trijunction has a first-harmonic correspondence to three $\pi$ junctions.
        (b) Energy landscape at $\phi_1=0$, with the two opposite-chirality minima marked by white circles.
        (c) Bond correlations $|C_1|$ and $C_3$, and local chirality variance $\langle\kappa^2\rangle$.
        Solid curves show thermal averages, while dashed curves show ground-state averages.
        The $x$-axis shows $T_{\mathrm{eff}}=T$ for thermal curves and $k_BT_{\mathrm{eff}}=\hbar\omega_p/2$ for quantum curves.
    }
    \label{fig:trijunction}
\end{figure}

The singlet-doublet transition in a quantum dot Josephson junction reverses its sine-like supercurrent; \emph{i.e.}, it shifts the energy-phase relation of the Josephson junction by $\pi$~\cite{van_dam_supercurrent_2006}.
If the quantum dot is coupled to more than two superconducting leads, each pair of leads contributes a $\pi$-junction-like energy-phase relation (EPR), as illustrated in Fig.~\ref{fig:trijunction}(a).
We neglect higher harmonics in the EPR -- which are much less pronounced in odd-parity junctions-- so that a multiterminal potential can be approximated as a sum of pairwise $\pi$-junction contributions:
\begin{equation}
    U(\{\phi_i\})=J\sum_{\langle i,j\rangle}\cos(\phi_i-\phi_j)~,\quad J>0~,
    \label{eq:rotor-potential}
\end{equation}
where $\phi_i$ is the superconducting phase of island $i$, $J$ is the Josephson coupling between each pair of leads, and the sum runs over the three pairs.

In a trijunction, the resulting EPR coincides exactly with the triangular $\pi$-junction loop shown in Fig.~\ref{fig:trijunction}(a)~\cite{teitel_josephson-junction_1983}, and is also identical to the first-harmonic approximation of a frustrated triangular XY antiferromagnet.
We show the EPR in Fig.~\ref{fig:trijunction}(b) and highlight with two degenerate minima for which the superconducting phases differ by $2\pi/3$ across each pair of leads, spontaneously forming a discrete vortex~\cite{feinberg_spontaneous_2014,manesco_loopless_2026}.
The two degenerate minima differ by the chirality
\begin{equation}
    \kappa = \frac{2}{3\sqrt3}\sum_{\circlearrowleft}\sin(\phi_i-\phi_j)~,
\end{equation}
of the discrete vortex.

The frustrated minima lock neighboring superconducting phases at differences of $\pm2\pi/3$.
Averaging with equal weights over the two opposite chiralities therefore gives $\langle e^{i(\phi_i-\phi_j)}\rangle=-1/2$.
Yet tripling the phase differences removes the mismatch entirely:
\begin{equation}
    e^{i3(\phi_i-\phi_j)}=1
\end{equation}
in either minimum, as illustrated in Fig.~\ref{fig:intro}(c).
This raises the question of whether correlations of the tripled phase can persist when ordinary phase correlations are suppressed.
To investigate this possibility, we introduce
\begin{equation}
    C_n^{ij} = \langle e^{in(\phi_i - \phi_j)}\rangle~,
\end{equation}
for an arbitrary pair of superconducting islands $i$ and $j$, and compare $C_1$ with $C_3$ under thermal and quantum fluctuations.
Here $\langle\cdots\rangle$ denotes a classical thermal average with Boltzmann weight $e^{-U/k_BT}$, or a quantum ground-state expectation value.
Writing the superconducting order parameter on island $i$ as $\Psi_{2e,i}\propto e^{i\phi_i}$, we recognize $C_1$ as its phase correlator.
Similarly, $C_3$ measures phase correlations of the composite field $\Psi_{2e,i}^{3}$, which carries charge $6e$.

We now compute $C_1$ and $C_3$ as a function of both temperature and charging energy using Eq.~\ref{eq:rotor-potential}.
The numerical thermal correlators in Fig.~\ref{fig:trijunction}(c) cross at $k_BT_\times/J\approx0.107$, where $C_3=|C_1|$.
The quantum ground-state correlators show a qualitatively similar crossover as the zero-point energy $\hbar\omega_p/2$ increases, where $\omega_p=\sqrt{12E_CJ}/\hbar$ is the plasma frequency of the trijunction and $E_C$ is the island charging energy.
This similarity is expected since, within the harmonic approximation, the variance of the phase fluctuations is identical for the quantum and classical thermal cases.
For the spinless network calculations below, we assume large capacitances and treat the superconducting phases classically.
We expect quantum phase fluctuations to affect the correlators $C_1$ and $C_3$ similarly to the thermal fluctuations considered below.

\section{Composite coherence in trijunction networks}

We now investigate whether the phases of a network of superconducting islands coupled via odd-parity quantum dots can exhibit long-distance $6e$ coherence.
We start with the classical regime, corresponding to large island capacitances, and neglect self-inductance effects to understand the interplay between phase frustration and thermal fluctuations.
As we estimate in Sec.~\ref{sec:experimental-considerations}, the large capacitances required to minimize quantum fluctuations result in a large separation between quantum dots.
For this reason, we also neglect coherent transfer of electrons between quantum dots via processes such as elastic cotunneling or crossed Andreev reflection.

\subsection{Third-harmonic Josephson response}

\begin{figure}[!t]
    \centering
    \includegraphics[width=\linewidth]{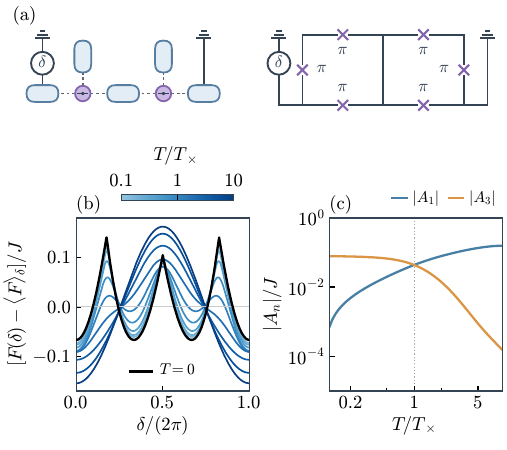}
    \caption{
        Threefold Josephson response of two frustrated trijunctions.
        (a) Odd-parity two-dot device and analog $\pi$-junction circuit highlighting the phase bias $\delta$, while the internal phases float.
        (b) Free energy of the Josephson device as a function of the imposed phase bias $\delta$ at different temperatures.
        (c) Temperature dependence of the first and third-harmonic responses of the free energy.
        The dotted line in panel (c) indicates the numerically determined crossover temperature $T_\times$.
    }
    \label{fig:double-dot-JJ}
\end{figure}

We start with a minimal example shown in Fig.~\ref{fig:double-dot-JJ}(a): a Josephson junction formed by two trijunctions in series with three floating superconducting islands.
Again, neglecting higher harmonics, we can map this device to an array of $\pi$-junctions also shown in Fig.~\ref{fig:double-dot-JJ}(a).
We compute the EPR of this device by minimizing the energy for each bias phase $\delta$.
We see that at zero temperature the EPR has three minima, a consequence of a pronounced third harmonic, as shown in Fig.~\ref{fig:double-dot-JJ}(b).
This is the behavior we would expect to see from a Josephson element carrying $6e$ supercurrent.

Similarly to a single trijunction, the $6e$ effects observed in a double-trijunction Josephson element are suppressed by thermal fluctuations.
In Fig.~\ref{fig:double-dot-JJ}(b) we show how the EPR changes with temperature, with a clear crossover from a triple-minima EPR to the usual cosine-like dependence on $\delta$.
We quantify this evolution in Fig.~\ref{fig:double-dot-JJ}(c), where we show the temperature dependence of the first and third harmonics of the EPR.
The numerically evaluated free-energy harmonics cross at $k_BT_\times/J\approx0.111$, where $|A_1|=|A_3|$.
Here, $A_n$ is the coefficient of $\cos(n\delta)$ in the equilibrium free energy.

The third harmonic can be enhanced relative to the first by connecting more trijunctions in series.
This relative enhancement is exponential in chain length when the variation of the free energy with the imposed phase is small compared with $k_BT$ (see App.~\ref{app:chain-harmonics}).

\subsection{Collective $6e$ coherence}

\begin{figure}[!t]
    \centering
    \includegraphics[width=\linewidth]{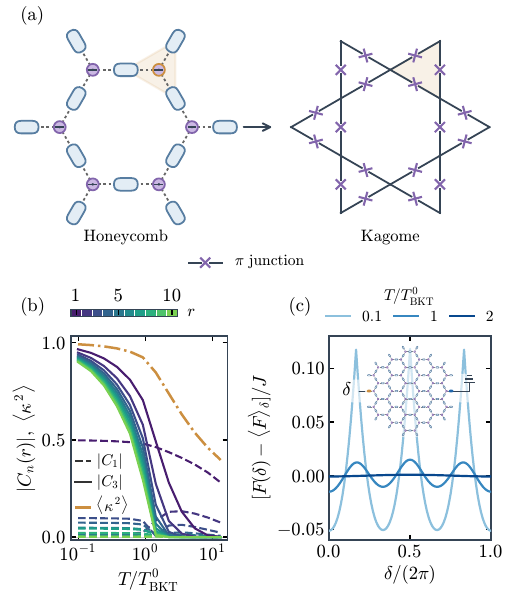}
    \caption{
        Thermal phase correlations, local chirality, and phase-biased response of trijunction arrays.
        (a) Honeycomb device and its Kagome $\pi$-junction analog, with the highlighted area showing the correspondence illustrated in Fig.~\ref{fig:trijunction}(a).
        (b) Correlation magnitudes $|C_1(r)|$ and $|C_3(r)|$ for the infinite Kagome XY antiferromagnet versus temperature, with colors indicating graph distances $r$.
        Temperatures are normalized by the estimate $T_{\mathrm{BKT}}^0$ defined in App.~\ref{app:temperatures}.
        The orange dash-dotted curve shows the local chirality variance $\langle\kappa^2\rangle$.
        (c) Phase-dependent free energy of a 54-trijunction array with open boundaries, obtained by tensor contraction (App.~\ref{tensor-networks}).
        Inset: scheme of the device with imposed phase bias $\delta$.
    }
    \label{fig:honeycomb}
\end{figure}

In the chain shown in Fig.~\ref{fig:double-dot-JJ}(a), some superconducting islands connect to only one trijunction, allowing their phases to adjust without constraints from neighboring trijunctions.
In a two-dimensional lattice, however, the geometry imposes additional constraints on the phase configurations.
We address how these constrained geometries affect the long-distance $6e$ coherence by considering the honeycomb superlattice shown in Fig.~\ref{fig:honeycomb}(a).
Notably, from the correspondence between the odd-parity trijunction and a triangle of $\pi$-shifted Josephson junctions, we can map this lattice exactly into an XY Kagome antiferromagnet, as shown in Fig.~\ref{fig:honeycomb}(a).

A Kagome XY antiferromagnet has a frustrated ground state, resulting in a highly degenerate manifold.
The states in this manifold correspond to all allowed chirality combinations for which the superconducting phases differ by $2\pi/3$ across each pair of neighboring islands.
Therefore, over long distances we should expect a random chirality configuration such that $C_1$ averages to zero.
At the same time, $C_3=1$ at zero temperature for any pair of superconducting phases.

At finite temperatures, all the correlations are suppressed, but at different rates.
As shown in Refs.~\cite{Song2023,Song2025}, the $6e$ order emerges through a BKT transition.
Below $k_B T_{\mathrm{BKT}} / J \approx 0.075$, $C_1$ is exponentially suppressed over a graph distance $r$, while the decay of $C_3(r)$ is algebraic.
We illustrate these correlations by computing $C_1$ and $C_3$ as functions of temperature at different graph distances $r$, as shown in Fig.~\ref{fig:honeycomb}(b) (details in App.~\ref{tensor-networks}).
The $6e$ order clearly dominates at low temperatures with a weak decay with distance.
These results establish frustrated trijunction lattices as a platform for quantum simulation of higher-order superconductivity.

Similarly to the two-trijunction Josephson element, we can probe the $6e$ order in a finite honeycomb array by imposing a phase difference $\delta$ between two superconducting islands.
Differently from the 2-trijunction device, however, the honeycomb lattice has a large number of internal islands whose phases can adjust to minimize the energy.
In Fig.~\ref{fig:honeycomb}(c) we show the phase-dependent free energy of a finite 54-trijunction array with open boundaries.
We compute the free energy by imposing a phase difference $\delta$ between two superconducting islands, while averaging over the remaining internal phases, as described in App.~\ref{tensor-networks}.
We observe a pronounced third harmonic in the free energy below the BKT transition, which is suppressed at higher temperatures.

\section{From phase locking to spin entanglement}

\begin{figure}[!t]
    \centering
    \includegraphics[width=\linewidth]{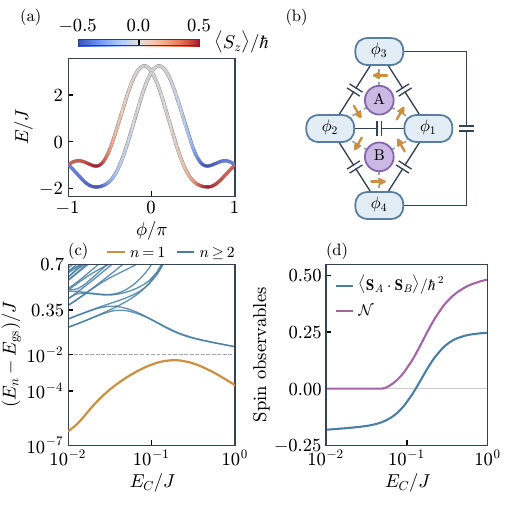}
    \caption{
        (a) Eigenvalues of $\hat W$ along $(\phi_1,\phi_2,\phi_3)=(\phi,0,-\phi)$ colored by their spin projection $\langle S_z\rangle/\hbar$.
        (b) Two-dot circuit with gold arrows indicating Rashba axes.
        (c) Excitation energies of the circuit.
        The $y$-axis scale is logarithmic below $10^{-2}J$ and linear above, with a dashed line marking the change.
        (d) Spin correlation  $\langle\mathbf S_A\cdot\mathbf S_B\rangle/\hbar^2$ and reduced-spin negativity $\mathcal N$.
        We added a $5\%$ imbalance to the charging energies to lift accidental degeneracies.
    }
    \label{fig:spin-dependence}
\end{figure}

In Sec.~\ref{sec:single-trijunction}, we showed that phase $6e$ coherence is suppressed by quantum fluctuations.
However, in the presence of spin-orbit coupling, the spin of the trapped quasiparticle depends on the phase configuration of the trijunction~\cite{padurariu_superconducting_2010,manesco_loopless_2026}.
We therefore explore how these phase fluctuations affect the spin configuration of trijunctions sharing superconducting islands.
We introduce the degeneracy-breaking effects of spin-orbit coupling (SOC) on the trijunction potential~\cite{manesco_loopless_2026} (the full expression is given in App.~\ref{app:spin-hamiltonian}):
\begin{equation}
    \hat W(\{\phi_i\}) = U(\{\phi_i\}) + \vec{V}(\{\phi_i\})\cdot\vec{\sigma}~,
    \label{eq:spin-potential}
\end{equation}
where $\vec V$ describes the phase-dependent spin-orbit coupling and $\vec\sigma$ is the vector of Pauli matrices.
The coupling strength is controlled by the spin precession angle $\alpha$ between the superconducting leads and the quantum dots.
The relative scalar and spin-dependent couplings determine which phase-locked spin configurations minimize the energy.
For Fig.~\ref{fig:spin-dependence}, we use the leading weak-tunnelling couplings of a symmetric two-orbital dot with level energies $(0,\epsilon)$, fixed odd parity, $\epsilon=2\Delta$, and $\alpha=\pi/6$, where $\Delta$ is the superconducting gap.
The isolated trijunction then has a time-reversed pair of spin-dependent winding minima shown in Fig.~\ref{fig:spin-dependence}(a).

The spin-chirality locking of the spin-energy phase relation in a trijunction therefore leads to a one-to-one correspondence between the classical chirality configurations found in the previous section and the spin configurations of trijunction lattices.
As $E_C$ grows, however, quantum fluctuations delocalize the phases, breaking the correspondence between classical chirality and spin configurations.
Nonetheless, the trapped spins remain correlated via the shared superconducting phases.
To uncover what happens in this regime, we consider the double-dot device shown in Fig.~\ref{fig:spin-dependence}(b).
In this configuration, the two quantum dots share two superconducting islands, and as a consequence, in the classical limit their chiralities are locked in an antiparallel configuration.
As a result, the spins of the trapped quasiparticles in the two dots are also locked in an antiparallel configuration.
Unlike the inductive coupling between spinmons, which vanishes at zero applied flux~\cite{tjernshaugen_coherent_2026}, the shared superconducting phases in our circuit mediate spin coupling and entanglement without an applied flux.

At small $E_C$, the two lowest states of the two-dot device are the symmetric and antisymmetric superpositions of the classical minima:
\begin{equation}
    |\Psi_\pm\rangle \sim \ket{\circlearrowleft, \uparrow}\otimes\ket{\circlearrowright,\downarrow} \pm \ket{\circlearrowright, \downarrow}\otimes\ket{\circlearrowleft,\uparrow}~,
\end{equation}
where $\ket{\circlearrowleft}$ and $\ket{\circlearrowright}$ are the two opposite winding phase packets of the two-trijunction device, and $\ket{\uparrow}$ and $\ket{\downarrow}$ indicate the spin configurations of the two dots.
As $E_C$ increases, the ground-state wavefunctions delocalize, while opposite-$z$ spin correlations persist and positive transverse correlations develop, resulting in a reduced-spin density matrix $\rho_{AB}$ with a high degree of entanglement.
We show this entanglement in in Fig.~\ref{fig:spin-dependence}(d) where we plot the spin correlation $\langle\mathbf S_A\cdot\mathbf S_B\rangle/\hbar^2$ and the reduced-spin negativity~\cite{vidal_computable_2002},
\begin{equation}
    \mathcal{N} = \frac{\|\rho_{AB}^{T_B}\|_1 - 1}{2}~,
\end{equation}
as a function of $E_C/J$.
The negativity $\mathcal{N}$ is zero for separable states and reach $1/2$ for maximally entangled pairs.
Here, $T_B$ denotes the partial transpose with respect to spin $B$, and $\|\cdot\|_1$ is the trace norm.
As $E_C$ increases the reduced-spin negativity approaches the Bell-state value of $1/2$.
Due to the computational cost, we leave an analysis of quantum phases in spinful lattices to future work.

\section{Beyond trijunctions}

\begin{figure}[!t]
    \centering
    \includegraphics[width=\linewidth]{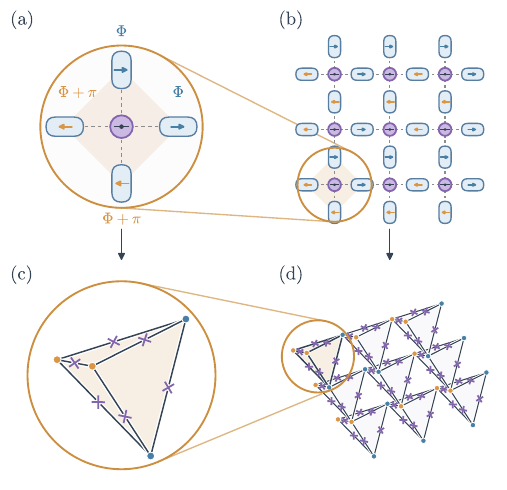}
    \caption{
        (a) Four-terminal junction in one collinear minimum, with two islands at phase $\Phi$ (blue) and two at $\Phi+\pi$ (orange).
        (b) Shared-island array: a square array in one compatible square-ice configuration; arrows indicate superconducting phases.
        (c) Tetrahedral $\pi$ circuit: the equivalent first-harmonic tetrahedron of six $\pi$ junctions.
        (d) Planar pyrochlore: the corresponding network, shown as a buckled layer of corner-sharing tetrahedra.
        The same phase colors are used throughout. The illustrated configuration is one member of the degenerate manifold, not a unique ordered state.
        Gold highlights and zoom lines identify corresponding local elements.
    }
    \label{fig:four-terminal}
\end{figure}

We now briefly discuss the generalization of these results to multiterminal junctions with more than three leads.
While trijunctions enforce a phase winding around the quantum dot, junctions with larger number of leads can enforce more complex phase configurations.
For example, in the 4-terminal junctions schematically shown in Fig.~\ref{fig:four-terminal}(a), it is possible to arrange the phases in two independent $\pi$-junction pairs, resulting in a continuum of degenerate minima along the line $(0,\pi,\theta,\theta+\pi)$, where $\theta$ is a free parameter.
Because all islands are connected in a 4-terminal junction, they map directly onto a tetrahedron of $\pi$ junctions, as shown in Fig.~\ref{fig:four-terminal}(c), for which these of degenerate minima were proposed as a resource for qubit protection~\cite{feigelman_superconducting_2004}.

Connecting these tetrajunctions in a square array with shared superconducting islands, as shown in Fig.~\ref{fig:four-terminal}(b), maps, in the first-harmonic approximation, onto the planar pyrochlore, or checkerboard, XY antiferromagnet in Fig.~\ref{fig:four-terminal}(d).
Thus, multiterminal junctions with more than three leads map to Josephson arrays with no planar two-terminal analog, and can be used to explore the physics of frustrated XY antiferromagnets in higher dimensions.
In these arrays, configurations with two phases $\theta$ and two phases $\theta+\pi$ on each tetrahedron obey the square-ice rule, giving an exponentially large number of compatible configurations~\cite{Lieb1967}.
At finite temperatures, thermal fluctuations favor a common phase, while conventional Néel order remains suppressed~\cite{Canals2004}.
The simulations of Ref.~\cite{Canals2004} find algebraic nematic correlations below an estimated BKT transition at $k_BT_{\mathrm{BKT}}/J\approx0.014$.
Since doubling the phases removes the $\pi$ mismatch, the nematic field corresponds to a $4e$ composite phase coherence, $\Psi_{4e}\propto\Psi_{2e}^2$.
This coexistence of composite phase coherence and many compatible phase configurations provides a route to high entropy superconductivity.
We leave the effects of the full microscopic potential and quantum charge fluctuations on this phase to future work.

\section{Experimental considerations}
\label{sec:experimental-considerations}

The critical temperature for the BKT transition in the honeycomb lattice is $k_B T_{\mathrm{BKT}} \sim 0.075 J$.
In InAs--Al hybrids, experiments have observed $J/h\sim \qty{2.5}{\giga\hertz}$~\cite{pita-vidal_strong_2024}, therefore resulting in $T_{\mathrm{BKT}} \sim \qty{10}{\m\kelvin}$, which is a challenging target for current dilution fridges.
An alternative way to achieve higher transition temperatures is to increase the Josephson coupling $J$ by using superconductors with larger gaps, such as Pb or Nb.
Hard induced gaps of $\Delta\simeq\qty{1.25}{\milli\electronvolt}$ in InAs--Pb~\cite{Kanne2021}, up to $\qty{1.3}{\milli\electronvolt}$ in PbTe--Pb~\cite{Gao2024}, and approximately $\qty{1.4}{\milli\electronvolt}$ in InSb--Pb~\cite{Li2024} have been reported; Josephson-junction transport also yields an induced gap of $\qty{1.3}{\milli\electronvolt}$ in InAsSb--Nb~\cite{Telkamp2026}, around 5 to 7 times larger than the typical InAs--Al gap $\sim\qty{0.2}{\milli\electronvolt}$.
In the intermediate-coupling regime, we expect that $J\propto\Delta$ at comparable dimensionless junction parameters; this scaling motivates a projected coupling $J/h\sim\qty{15}{\giga\hertz}$, giving $T_{\mathrm{BKT}}\sim\qty{50}{\m\kelvin}$.

As shown in Fig.~\ref{fig:trijunction}, large capacitances are required to suppress quantum fluctuations that destroy the $6e$ order.
At the projected coupling $J/h\sim\qty{15}{\giga\hertz}$, a charging energy $E_C/h\sim\qty{100}{\mega\hertz}$ gives a zero-point fluctuation scale $\hbar\omega_p/(2J)\sim0.14$, close to the local correlation crossover discussed above.
Assuming a dielectric thickness of $\qty{10}{\nano\meter}$ and a dielectric constant $\epsilon_r\sim 10$, we estimate that the area of the superconducting islands should be $\gtrsim \qty{25}{\micro\meter\squared}$, corresponding to a length scale of $\gtrsim \qty{5}{\micro\meter}$.
Since this length scale is much larger than the coherence length of the superconducting leads, we assume that the effects of elastic cotunneling and crossed Andreev reflection are suppressed.

Another natural question is whether the effects we reported are robust to disorder in the Josephson couplings.
There are two main sources of disorder in the Josephson couplings: (i) a random distribution of the tunneling amplitudes between the quantum dot and the superconducting leads, and (ii) a random distribution of the energy levels in the quantum dot.
While disorder in the chemical potential does not move the position of the minima in the energy-phase relation, it changes the Josephson couplings $J$, and can therefore affect the critical temperature of the BKT transition.
On the other hand, disorder in the tunneling amplitudes breaks the local threefold rotational symmetry of the lattice, moving the position of the minima in the energy-phase relation.
Similarly to the temperature and charging effects, we estimate the disorder scale by comparing $\sigma_J^2/J$ with $k_BT_{\mathrm{BKT}}$, where $\sigma_J$ is the rms disorder in the effective Josephson couplings.
The harmonic estimate gives $\sigma_J\sim\sqrt{2Jk_BT_{\mathrm{BKT}}/3}\approx0.22J$ (see App.~\ref{app:disorder} for details).

\section{Conclusion}

In this work, we showed how filling states in quantum dots frustrates the superconducting order parameter in multiterminal Josephson networks without explicitly breaking time-reversal symmetry.
As we tile these frustrated elements into arrays, their shared superconducting phases mediate long-range $6e$ phase coherence, while the usual $2e$ order is suppressed.
Our calculations and estimates indicate robustness of $6e$ coherence against thermal and quantum fluctuations and disorder.
The threefold phase structure produces a pronounced third-harmonic Josephson response, which can be probed in transport measurements both in linear quantum dot chains and in two-dimensional arrays.
While quantum fluctuations suppress the $6e$ order, spin-orbit coupling locks the spin of the trapped fermions to the phase arrangement of the superconducting islands.
As a result, the shared superconducting islands mediate spin correlations and entanglement between distant quantum dots, generating a long-range antiferromagnetic interaction and spin entanglement.
Our results therefore establish fermionic frustration in multiterminal Josephson networks as a resource for high entropy superconductivity and inter-dot spin correlations and entanglement.

\begin{acknowledgements}

The authors acknowledge Frederik Philipsen, Brian Andersen, Jens Paaske, and Mats Barkman for useful discussions.
This research was supported by the European Research Council (Grant Agreement No. 856526).

\emph{AI usage statement.} We used OpenAI Codex (versions 5.6-6.1) to produce code and figures, develop analytical derivations, and review manuscript.
We directed this work and reviewed the resulting code, figures, derivations, and text.

\emph{Code availability.} The code and data used to produce the results presented in this work are available on Zenodo~\cite{rigotti_manesco_zenodo_2026}.

\end{acknowledgements}

\bibliography{refs}

\appendix
\setcounter{secnumdepth}{1}

\section{Crossover and transition temperatures}
\label{app:temperatures}

\subsection{Single trijunction}

The chain has $N$ dots, each forming a trijunction.
For $N=1$, fixing $\phi_1=0$ gives the energy
\begin{equation}
 U=J[\cos\phi_2+\cos\phi_3+\cos(\phi_2-\phi_3)].
\end{equation}
The partition function is
\begin{equation}
 Z_1=\int_0^{2\pi}\frac{d\phi_2\,d\phi_3}{(2\pi)^2}e^{-U/(k_BT)}.
\end{equation}
The thermal average includes both chiral minima. The bond correlations are
\begin{equation}
 C_n=\frac13\sum_{i<j}\langle\cos(n\phi_{ij})\rangle_T,
 \qquad \phi_{ij}=\phi_i-\phi_j.
 \label{eq:single-dot-bond-correlation}
\end{equation}
Time reversal makes the sine averages vanish. With equal couplings all bonds are
equivalent, so $C_n=C_n^{ij}$ as defined in the main text.
The crossing $C_3=|C_1|$ defines $T_{\times,1}$.

Near either chiral minimum, write $u_i=\phi_i-\phi_i^{\min}$. Then
\begin{equation}
 U=-\frac{3J}{2}+\frac{J}{4}\sum_{i<j}(u_i-u_j)^2
   +O(u^3).
 \label{eq:single-dot-winding-expansion}
\end{equation}
Each bond angle fluctuates about $\pm2\pi/3$ by $u=u_i-u_j$.
At dimensionless temperature $\tau=k_BT/J$, equipartition for the two
relative modes of stiffness $3J/2$ gives $\langle u^2\rangle=4\tau/3$.
A Gaussian displacement reduces the $n$th harmonic by
$e^{-n^2\langle u^2\rangle/2}$, giving
\begin{equation}
 C_1\simeq-\frac12 e^{-2\tau/3},\qquad C_3\simeq e^{-6\tau}.
 \label{eq:single-dot-gaussian-correlations}
\end{equation}
The Gaussian crossing is
\begin{equation}
 \frac{k_BT_{\times,1}^{\mathrm{G}}}{J}
 =\frac{3\ln2}{16}\simeq0.130.
\end{equation}
The full periodic potential gives
\begin{equation}
 \frac{k_BT_{\times,1}}{J}\simeq0.107.
 \label{eq:single-crossing}
\end{equation}

For equal island charging energies, the two orthonormal modes perpendicular
to the uniform phase are independent oscillators. Their phase and charge
coordinates obey $[q_\nu,p_{\nu'}]=i\delta_{\nu\nu'}$ for $\nu,\nu'=1,2$,
with charges measured in Cooper pairs. Each mode has Hamiltonian
\begin{equation}
 H_\nu=4E_C p_\nu^2+\frac{3J}{4}q_\nu^2,
 \qquad \hbar\omega_p=\sqrt{12E_CJ}.
\end{equation}
Its ground-state phase variance equals the classical thermal variance at
$k_BT=\hbar\omega_p/2$. Figure~\ref{fig:trijunction} uses this equivalence
to plot both fluctuation scales in units of $k_BT_{\times,1}$.

\subsection{Double trijunction}

For one dot with end-to-end phase drop $\delta$, integrating over the
floating side phase $\phi$ gives the weight
\begin{equation}
 \begin{aligned}
 w(\delta)&=\int_0^{2\pi}\frac{d\phi}{2\pi}
 e^{-U(0,\delta,\phi)/(k_BT)}\\
 &=e^{-\cos\delta/\tau}
 I_0\!\left(\frac{2\cos(\delta/2)}{\tau}\right),
 \end{aligned}
 \label{eq:equal-dot-weight}
\end{equation}
where $I_0$ is the modified Bessel function.
Integrating over $\delta$ gives the single-dot partition function,
\begin{equation}
 Z_1=\int_0^{2\pi}\frac{d\delta}{2\pi}\,w(\delta).
\end{equation}

For $N=2$, fixing the total drop $\delta$ leaves one free phase $\theta$.
The drops across the two dots are $\theta$ and $\delta-\theta$, so their
weights enter the partition function through a circular convolution,
\begin{align}
 Z_2(\delta)&=\int_0^{2\pi}\frac{d\theta}{2\pi}
 w(\theta)w(\delta-\theta),\label{eq:two-dot-convolution}\\
 F_2(\delta)&=-k_BT\ln Z_2(\delta).
\end{align}
To evaluate the convolution in Eq.~\eqref{eq:two-dot-convolution},
we divide each weight by $Z_1$ so that it integrates to one. Its Fourier
coefficients are then thermal averages,
\begin{align*}
 \frac{w(\delta)}{Z_1}&=\sum_{n\in\mathbb Z} r_n(T)e^{in\delta},\\
 r_n(T)&=\int_0^{2\pi}\frac{d\delta}{2\pi}\,
            \frac{w(\delta)}{Z_1}e^{-in\delta}.
\end{align*}
Since $w$ is even, $r_n=\langle\cos(n\delta)\rangle_T$.
For equal couplings every bond has this correlation, so $r_n=C_n$
in Eq.~\eqref{eq:single-dot-bond-correlation}. Substituting the two
Fourier series gives
\begin{align*}
 \frac{Z_2(\delta)}{Z_1^2}
 &=\sum_{n,m\in\mathbb Z}r_n r_m e^{im\delta}
   \underbrace{\int_0^{2\pi}\frac{d\theta}{2\pi}\,e^{i(n-m)\theta}}_{\delta_{nm}}\\
 &=\sum_{n\in\mathbb Z}r_n^2 e^{in\delta}.
\end{align*}
Each dot supplies one factor of $r_n$, so the two-dot coefficient is
$p_{2,n}=r_n(T)^2$. The first index counts dots and the second labels the harmonic.
Normalization gives $r_0=1$, and evenness gives $r_{-n}=r_n$.
Thus,
\begin{equation}
 \frac{Z_2(\delta)}{Z_1^2}=1+2\sum_{n\geq1}p_{2,n}\cos(n\delta).
 \label{eq:two-dot-partition-harmonics}
\end{equation}
With the same index convention, the free-energy coefficients $A_{2,n}$
(written $A_n$ in the main text) satisfy
\begin{equation}
 F_2(\delta)-\langle F_2\rangle_\delta=\sum_{n\geq1}A_{2,n}(T)\cos(n\delta).
\end{equation}
When $Z_2$ varies weakly with phase, expanding the logarithm to first order
gives $A_{2,n}\simeq-2k_BT\,p_{2,n}$, so
\begin{equation}
 \left|\frac{A_{2,3}(T)}{A_{2,1}(T)}\right|
 \simeq\left|\frac{r_3(T)}{r_1(T)}\right|^2.
 \label{eq:two-dot-harmonic-ratio}
\end{equation}
The crossover $T_{\times,2}$ occurs at $|A_{2,3}|=|A_{2,1}|$, so this approximation
predicts it at $|r_3|=|r_1|$.

Since $r_n=C_n$ for equal couplings, the terms linear in $\tau$ follow
directly from Eq.~\eqref{eq:single-dot-gaussian-correlations}.
Retaining the cubic and quartic terms in the energy expansion gives the
coefficients of $\tau^2$.
\begin{equation}
 \begin{aligned}
 r_1&=-\frac12+\frac{\tau}{3}+\frac{2\tau^2}{9}+O(\tau^3),\\
 r_3&=1-6\tau+8\tau^2+O(\tau^3).
 \end{aligned}
\end{equation}
Near the crossing $r_1<0$ and $r_3>0$. Setting $r_3=-r_1$ in the
quadratic expansion gives $148\tau^2-102\tau+9=0$, with smaller root
\begin{equation}
 \frac{k_BT_{\times,2}^{\mathrm{an}}}{J}
 =\frac{51-\sqrt{1269}}{148}\simeq0.104.
\end{equation}

Numerically evaluating the convolution with the full periodic weight and
taking the logarithm before extracting $A_{2,1}$ and $A_{2,3}$ retains harmonic
mixing. The resulting crossing is
\[
 \frac{k_BT_{\times,2}}{J}\simeq0.111.
\]
The analytical estimate is about $6\%$ lower. The interpolated numerical
crossing sets the temperature scale and the dotted line in Fig.~\ref{fig:double-dot-JJ}.
For the current $I=(2e/\hbar)\partial_\delta F_2$, differentiating each
harmonic gives $I_{2,n}=-(2e/\hbar)nA_{2,n}$. The current harmonics therefore
cross at $|A_{2,1}|=3|A_{2,3}|$.

\subsection{Kagome BKT scale}

Fractional vortices in the Kagome XY model have phase winding $2\pi/3$.
Their BKT transition satisfies~\cite{Korshunov2002}
\begin{equation}
 k_BT_{\mathrm{BKT}}=\frac{\pi}{18}\rho_s(T_{\mathrm{BKT}}),
\end{equation}
where $\rho_s$ is the stiffness of the original phase field.
Using the zero-temperature stiffness $\rho_s(0)=\sqrt3J/4$~\cite{Korshunov2002}
gives
\begin{equation}
 \frac{k_BT_{\mathrm{BKT}}^0}{J}
 =\frac{\pi\sqrt3}{72}\simeq0.0756.
\end{equation}
This estimate sets the temperature unit in Fig.~\ref{fig:honeycomb}.
For the ideal array we use the thermodynamic-limit result
$k_BT_{\mathrm{BKT}}/J\simeq0.075$ from Ref.~\cite{Song2023}
and compute correlations at selected temperatures.

\section{Exact chain convolution and leading harmonics}
\label{app:chain-harmonics}

We extend the two-dot calculation of Appendix~\ref{app:temperatures} to a
chain of $N$ dots. Each dot $i$ can have different positive Josephson
couplings $a_i,b_i,c_i$. On dot $i$, let $\delta_i$ and $\theta_i$ be the
phases of the right and floating terminals, respectively, both measured
relative to its left terminal. The energy of that dot is
\begin{equation}
 H_i=a_i\cos\delta_i+b_i\cos\theta_i+c_i\cos(\theta_i-\delta_i).
\end{equation}
The two terms containing $\theta_i$ combine into a single cosine of amplitude
$|b_i+c_i e^{i\delta_i}|$. Integrating over $\theta_i$ with measure
$d\theta_i/(2\pi)$ as in Eq.~\eqref{eq:equal-dot-weight} gives
\begin{equation}
 w_i(\delta_i)=e^{-a_i\cos\delta_i/(k_BT)}
 I_0\!\left(\frac{|b_i+c_i e^{i\delta_i}|}{k_BT}\right).
 \label{eq:disorder-local-weight}
\end{equation}
Each dot has its own normalization $\langle w_i\rangle_{\delta_i}$,
where the brackets denote a uniform average over the indicated phase. Its weight is even,
so its normalized Fourier coefficients are
\begin{equation}
 r_{i,n}=\frac{\langle w_i(\delta_i)\cos(n\delta_i)\rangle_{\delta_i}}
              {\langle w_i\rangle_{\delta_i}}.
\end{equation}
Thus $r_{i,n}$ measures the correlation across the endpoint bond of dot $i$.
For $a_i=b_i=c_i=J$, all three bonds are equivalent and $r_{i,n}=C_n$.

Imposing a phase difference $\delta$ between the ends of the chain fixes
the sum of the local phase drops,
\[
 \sum_{i=1}^N\delta_i \equiv \delta \pmod{2\pi}.
\]
In Eq.~\eqref{eq:two-dot-partition-harmonics}, the coefficient $p_{2,n}=r_n^2$ comes
from one factor per dot. For two different dots the same shared-phase
integral gives $r_{1,n}r_{2,n}$. Each additional dot contributes its
coefficient at the same order $n$, giving
\begin{align}
 \frac{Z_N(\delta)}{\langle Z_N\rangle_\delta}
 &=1+2\sum_{n\geq1}p_{N,n}\cos(n\delta),
 \label{eq:disorder-chain-partition}\\
 p_{N,n}&=\prod_{i=1}^N r_{i,n}.
\end{align}
Here $\langle Z_N\rangle_\delta=\prod_i\langle w_i\rangle_{\delta_i}$
plays the role of $Z_1^2$ in the two-dot result. To obtain the free-energy
harmonics, we write
\begin{equation}
 F_N(\delta)-\langle F_N\rangle_\delta=\sum_{n\geq1}A_{N,n}\cos(n\delta).
\end{equation}
The product rule is exact for $Z_N$. When $Z_N$ varies weakly with
phase, expanding $F_N=-k_BT\ln Z_N$ to first order gives
\begin{equation}
 A_{N,n}\simeq-2k_BT\,p_{N,n}.
 \label{eq:disorder-chain-leading}
\end{equation}
For identical dots, $p_{N,n}=r_n(T)^N$. The two-dot harmonic ratio
in Eq.~\eqref{eq:two-dot-harmonic-ratio} then generalizes to
\begin{equation}
    \left|\frac{A_{N,3}}{A_{N,1}}\right|
    \simeq\left|\frac{r_3(T)}{r_1(T)}\right|^N.
\end{equation}
In this approximation the third-to-first harmonic ratio grows exponentially
with $N$ when $|r_3|>|r_1|$. More generally, the third harmonic decays
more slowly than harmonic $n$ if $|r_3|>|r_n|$. At finite temperature
both amplitudes vanish in the long-chain limit.

\section{Disorder in chains and the Kagome lattice}
\label{app:disorder}

We study static variations of the effective Josephson couplings in the first-harmonic phase model, with mean $J>0$.
These variations preserve global $U(1)$ symmetry and time reversal but can break the equivalence of the contacts.
We first consider finite chains, then estimate the lattice response around a fixed ground state of the clean array.
Throughout this appendix, an overbar denotes the disorder ensemble average; thermal averages retain angle brackets.

\subsection{Third-harmonic dominance in disordered chains}

\begin{figure}[t]
 \centering
 \includegraphics[width=\columnwidth]{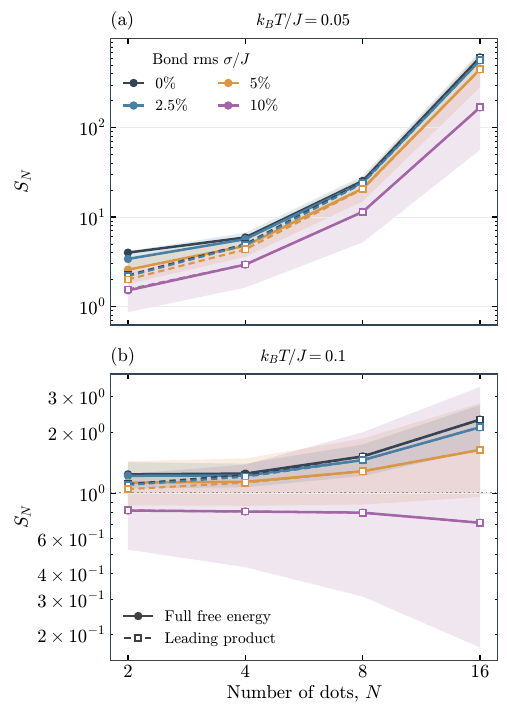}
 \caption{Relative third-harmonic strength in finite disordered trijunction chains at (a) $k_BT/J=0.05$ and (b) $0.10$.
 Colors denote rms effective-bond disorder.
 We use 65,536 realizations per nonzero disorder strength.
 Solid curves with circles show the disorder median of $S_N$; shading gives the central 68\% of individual realizations.
 Dashed curves with open squares show the leading estimate, Eq.~\eqref{eq:disorder-chain-leading}, for the same ensemble without fitted parameters.
 The dotted line marks $S_N=1$.}
 \label{fig:chain-disorder}
\end{figure}

We compute the chain free energy using the convolution in Appendix~\ref{app:chain-harmonics}.
Each bond is drawn independently from a uniform distribution centered at $J$, with half-width $\sqrt3\sigma$.
Here $\sigma$ is the rms effective-bond disorder $\sigma_J$ of the main text.
All bonds remain positive for the strengths in Fig.~\ref{fig:chain-disorder}.
We measure the relative third-harmonic strength by
\begin{equation}
 S_N=\frac{|A_{N,3}|}
 {\max_{m\in\{1,2,4,5\}}|A_{N,m}|}.
 \label{eq:disorder-selectivity}
\end{equation}
The sixth harmonic respects $2\pi/3$ periodicity and is excluded from the denominator.

In the leading approximation of Eq.~\eqref{eq:disorder-chain-leading}, each dot contributes a factor $|r_{i,3}/r_{i,m}|$ to the ratio $|A_{N,3}/A_{N,m}|$.
The largest competing amplitude $|A_{N,m}|$ gives the smallest such ratio. Taking logarithms gives
\begin{equation}
 \ln S_N^{\mathrm{lead}}
 =\min_{m\in\{1,2,4,5\}}\sum_{i=1}^N
 \ln\left|\frac{r_{i,3}}{r_{i,m}}\right|.
\end{equation}
For independent, identically distributed elements with finite absolute logarithmic moments, the law of large numbers gives
\begin{equation}
 \frac{\ln S_N^{\mathrm{lead}}}{N}\longrightarrow
 \min_{m\in\{1,2,4,5\}}
 \overline{\ln|r_3/r_m|}.
 \label{eq:disorder-selectivity-rate}
\end{equation}
Within this approximation, $S_N$ grows exponentially with chain length when the limit in Eq.~\eqref{eq:disorder-selectivity-rate} is positive and decreases when it is negative.
For coupling variations with zero mean, the linear correction to this growth rate averages to zero.
The leading change is of order $(\sigma/J)^2$ when the local Fourier coefficients are smooth and nonzero, so a positive clean rate survives sufficiently weak disorder.

Taking the logarithm of $Z_N$ mixes harmonics, an effect that matters for short chains and is absent from the leading approximation.
A larger $S_N$ means a stronger third harmonic relative to its competitors; its absolute amplitude can still decrease.

\subsection{Disorder in the Kagome lattice}

To estimate the phase distortions caused by weak bond disorder, we consider the Hamiltonian
\begin{equation}
 H=\sum_{\langle ij\rangle}(J+\delta J_{ij})\cos(\phi_i-\phi_j).
\end{equation}
The sum runs over nearest-neighbor bonds. We choose a ground state $\phi_i^0$ of the clean array independently of the disorder and let $u_i=\phi_i-\phi_i^0$.
On bond $ij$, the clean phase difference $\phi_{ij}^0=\phi_i^0-\phi_j^0$ is $\pm 2\pi/3$ modulo $2\pi$.
Changing this angle by $u_{ij}=u_i-u_j$ changes the bond energy by
\begin{equation}
 \Delta H_{ij}\simeq
 -J\sin\phi_{ij}^0\,u_{ij}+\frac J4 u_{ij}^2
 -\delta J_{ij}\sin\phi_{ij}^0\,u_{ij}.
\end{equation}
We keep the clean energy to second order in $u_{ij}$ and the disorder contribution to first order in $u_{ij}$.

In a clean triangle, an island at phase $\phi$ has neighbors at $\phi\pm2\pi/3$, so their torques cancel.
Unequal bond strengths leave a net torque $f_i$. For a fixed realization of the bonds, the zero-temperature displacement $u_i^{(0)}$ balances this torque against the restoring torque from the neighboring bonds,
\begin{equation}
 \frac{J}{2}\sum_{j\sim i}(u_i^{(0)}-u_j^{(0)})=f_i,\qquad
 f_i=\sum_{j\sim i}\delta J_{ij}\sin\phi_{ij}^0.
 \label{eq:disorder-lattice-balance}
\end{equation}
The sum over neighbors $j\sim i$ on the left is the discrete Laplacian $(Lu^{(0)})_i$.
A uniform shift leaves all bond differences unchanged, so we fix the mean displacement to zero. On this subspace the inverse of $L$ is the lattice Green function $G$.

At fixed disordered couplings, the equilibrium displacement and the energy change relative to the unrelaxed phases are
\begin{equation}
 u_i^{(0)}=\frac{2}{J}\sum_j G_{ij}f_j,\qquad
 \Delta E=-\frac12\sum_i f_i u_i^{(0)}.
 \label{eq:disorder-lattice-response}
\end{equation}
To find $\Delta E$, multiply the torque balance by $u_i^{(0)}$ and sum over sites. The two ends of each bond together give $(u_i^{(0)}-u_j^{(0)})^2$, so the elastic energy is $\tfrac12\sum_i f_i u_i^{(0)}$. The disorder term contributes $-\sum_i f_i u_i^{(0)}$.

Suppose the bond variations are independent, with zero mean and variance $\sigma^2$.
Each bond exerts opposite torques on its ends. It therefore adds $\sigma^2\sin^2\phi_{ij}^0=3\sigma^2/4$ to the torque variance at each end and $-3\sigma^2/4$ to the covariance between them.
Summing these independent bond contributions gives $\overline{f_i f_j}=(3\sigma^2/4)L_{ij}$.
For one realization, multiplying the equilibrium displacements at sites $i$ and $j$ from Eq.~\eqref{eq:disorder-lattice-response} gives
\[
 u_i^{(0)}u_j^{(0)}=\frac{4}{J^2}
 \left(\sum_k G_{ik}f_k\right)
 \left(\sum_\ell G_{j\ell}f_\ell\right).
\]
The clean lattice fixes $G$, so only the torque product $f_k f_\ell$ needs averaging.
\begin{equation}
 \begin{aligned}
 \overline{u_i^{(0)}u_j^{(0)}}
 &=\frac{4}{J^2}\sum_{k,\ell}G_{ik}\,\overline{f_k f_\ell}\,G_{j\ell}\\
 &=\frac{3\sigma^2}{J^2}(GLG)_{ij}
 =\frac{3\sigma^2}{J^2}G_{ij}.
 \end{aligned}
 \label{eq:disorder-lattice-covariance}
\end{equation}
The torque covariance supplies $L$ between the two factors of the symmetric Green function $G$. Since $LG$ is the identity on zero-mean displacements, $GLG=G$.

At finite temperature, write $u_i=u_i^{(0)}+\delta u_i$, where $\delta u_i$ is the thermal fluctuation at fixed disorder.
Equation~\eqref{eq:disorder-lattice-covariance} gives the covariance of the zero-temperature shifts, and the thermal covariance is $\langle\delta u_i\,\delta u_j\rangle=(2k_BT/J)G_{ij}$. Thus
\begin{equation}
 \begin{aligned}
 \overline{\langle(u_i-u_j)^2\rangle}
 &=\overline{\big\langle[(u_i^{(0)}-u_j^{(0)})+(\delta u_i-\delta u_j)]^2\big\rangle}\\
 &=\overline{(u_i^{(0)}-u_j^{(0)})^2}+\overline{\langle(\delta u_i-\delta u_j)^2\rangle}\\
 &=\frac{3\sigma^2}{J^2}R_{ij}+\frac{2k_BT}{J}R_{ij}.
 \end{aligned}
 \label{eq:disorder-lattice-variance}
\end{equation}
The cross term vanishes because $\langle\delta u_i-\delta u_j\rangle=0$ for each disorder realization.
Both variances contain $R_{ij}=G_{ii}+G_{jj}-2G_{ij}$, the effective resistance between sites $i$ and $j$ for unit bond resistances.
Tripling the clean phases makes them equal modulo $2\pi$, so the phase factor in $C_3^{ij}$ is $e^{3i(u_i-u_j)}$.
For Gaussian bond disorder, the harmonic displacement $u_i-u_j$ is Gaussian with zero mean under the combined thermal and disorder average. Applying the Gaussian average used in Eq.~\eqref{eq:single-dot-gaussian-correlations} with $n=3$ gives
\begin{equation}
 \overline{C_3^{ij}}=
 \exp\!\left[-\frac92\,\overline{\langle(u_i-u_j)^2\rangle}\right].
 \label{eq:disorder-lattice-correlation}
\end{equation}

The clean lattice has bare stiffness $\rho_s=\sqrt{3}J/4$~\cite{Korshunov2002}.
At large separation $r$, measured in unit cells, the two-dimensional Gaussian modes give a phase-difference variance of $(k_BT/\pi\rho_s)\ln r+O(1)$.
Matching this to the thermal term $(2k_BT/J)R(r)$ in Eq.~\eqref{eq:disorder-lattice-variance} gives
\begin{equation}
 R(r)=\frac{J}{2\pi\rho_s}\ln r+O(1)
     =\frac{2}{\sqrt3\pi}\ln r+O(1).
\end{equation}
Along a fixed direction, the graph distance in Fig.~\ref{fig:honeycomb}(b) differs by a constant scale, which changes the correlation prefactor but not its exponent.
Hence
\begin{equation}
 \overline{C_3(r)}\sim A r^{-\eta_3},\qquad
 \eta_3=\frac{9}{\sqrt3\pi}
 \left(\frac{2k_BT}{J}+\frac{3\sigma^2}{J^2}\right).
\end{equation}
Here $A$ is a nonuniversal amplitude set by the short-distance contribution to the phase variance.
The disorder contribution to the harmonic variance is therefore equivalent to the shift $k_BT\mapsto k_BT+3\sigma^2/(2J)$.
The relaxation energy in Eq.~\eqref{eq:disorder-lattice-response} can favor a different ground state of the clean array at order $\sigma^2/J$.
Within the harmonic approximation about a fixed clean ground state, third-harmonic correlations remain algebraic in the presence of weak bond disorder.
The decay exponent increases only quadratically with the disorder strength $\sigma/J$.

\section{Tensor-network calculation of Kagome correlations and free energy}
\label{tensor-networks}

We compute the correlations of the infinite Kagome XY antiferromagnet by expressing the partition function as a network of triangle tensors.
Every bond belongs to one triangle, whose Boltzmann weight and Fourier coefficients are
\begin{align}
 W(\boldsymbol\phi)&=e^{-\beta J(\sum_{i<j}\cos\phi_{ij}+3/2)},\\
 \widehat W_{\mathbf n}
 &=\int\frac{d^3\phi}{(2\pi)^3}W(\boldsymbol\phi)e^{-i\mathbf n\cdot\boldsymbol\phi}.
\end{align}
Here $\beta=1/(k_BT)$, $\phi_{ij}=\phi_i-\phi_j$, and $\mathbf n=(n_1,n_2,n_3)$ contains the integer Fourier indices.
Each phase is integrated over $[0,2\pi)$.
The constant energy shift keeps the weight bounded by one without changing the correlations.
Global phase invariance requires $n_1+n_2+n_3=0$, leaving two integrals that we evaluate by periodic quadrature.
We truncate the Fourier indices to $|n_i|\leq m$.

We combine two neighboring triangles into a transfer matrix product operator and approximate its dominant boundary state with a uniform matrix product state using MPSKit.jl~\cite{devos_mpskit_2026}.
We use the $U(1)$ charge sectors in TensorKit.jl~\cite{devos_tensorkit_2025} to preserve the symmetry without selecting a particular phase pattern or chirality.
The Fourier cutoff retains continuous phases.

We obtain $C_n(r)$ by inserting $e^{in\phi_i}$ and $e^{-in\phi_j}$ into the tensor contraction and dividing by the partition function.
The separation $r$ counts nearest-neighbor bonds along a fixed direction of the Kagome lattice.
Figure~\ref{fig:honeycomb}(b) uses Fourier cutoff $m=32$ and boundary bond dimension $\chi=40$.
The same triangle tensors describe the open 54-trijunction array in Fig.~\ref{fig:honeycomb}(c), where each bond carries the Josephson coupling of Eq.~\eqref{eq:rotor-potential}.
Fixing the phases at opposite contacts to $0$ and $\delta$ and integrating the remaining island phases over $[0,2\pi)$ gives the phase-dependent partition function $Z(\delta)$.
We contract the finite network with $m=48$ and $\chi=32$ to obtain $F(\delta)=-k_BT\ln Z(\delta)$, plotted in panel~(c) with its phase average subtracted.

\section{Minimal spin-orbit Hamiltonian}
\label{app:spin-hamiltonian}

We describe the two-dot circuit in Fig.~\ref{fig:spin-dependence} using the local spin-phase relation of Ref.~\cite{manesco_loopless_2026}.

\subsection{Local doublet potential}

We use a symmetric finite-contact model with two normal orbitals separated by $\epsilon>0$ and fixed odd electron parity.
The three superconducting islands of the trijunction have spin-singlet pairing with gap $\Delta$ and chemical potential $\mu_S=0$.
Each couples to the dot with the same tunnel amplitude $t$.
The three bonds from the dot to the islands lie $120^\circ$ apart in the plane.
For a bond pointing along $\widehat{\mathbf r}_i$, the Rashba axis is $\mathbf d_i=\widehat{\mathbf z}\times\widehat{\mathbf r}_i$, as shown by the gold arrows in Fig.~\ref{fig:spin-dependence}(b).
The tunnelling matrices for the two orbitals are
\begin{equation}
 T_{1i}=t e^{i\alpha\mathbf d_i\cdot\vec\sigma/2},
 \qquad
 T_{2i}=t e^{-i\alpha\mathbf d_i\cdot\vec\sigma/2}.
 \label{eq:spin-tunnelling}
\end{equation}
For $t\ll\Delta$ and $t^2/\Delta\ll\epsilon$, the leading odd-doublet Hamiltonian is $\hat W$ of Eq.~\eqref{eq:spin-potential}, up to a phase-independent constant.
Writing $\phi_{ij}=\phi_i-\phi_j$, its scalar and spin-dependent parts are
\begin{align}
 U&=E_0\sum_{i<j}\cos\phi_{ij},\\
 \vec V&=E_\sigma\sum_{i<j}\mathbf b_{ij}\sin\phi_{ij}.
\end{align}
The dimensionless pair vectors are
\begin{equation}
 \mathbf b_{ij}=\sin\alpha\cos\alpha(\mathbf d_i-\mathbf d_j)
 +\sin^2\alpha\,\mathbf d_i\times\mathbf d_j.
 \label{eq:spin-pair-vector}
\end{equation}
The scalar and spin-dependent terms have separate couplings.
With $x=\epsilon/\Delta$, this finite-contact model gives
\begin{align}
 E_0&=\frac{3t^4}{2\Delta^3}
 \left[1-\cos^2\alpha\frac{x+2}{x(x+1)^2}\right],\\
 E_\sigma&=\frac{2t^4}{\Delta^3x}.
 \label{eq:spin-couplings}
\end{align}
For $\epsilon=2\Delta$ and $\alpha=\pi/6$, we obtain $J=E_0=5E_\sigma/4$ and use $J$ as the energy unit.
The circuit uses this leading potential, with local branches $U_\pm=U\pm|\vec V|$.

\subsection{Two-dot circuit and numerical treatment}

Using the numbering in Fig.~\ref{fig:spin-dependence}, dots $A,B$ share islands 1 and 2, with separate outer islands 3 and 4, respectively.
We set the offset charges to zero and work in the sector of zero total charge.
The circuit Hamiltonian is
\begin{equation}
 H=H_C+\sum_{d=A,B}\hat W_d(\boldsymbol\phi_d),
\end{equation}
with charging term
\begin{equation}
 H_C=4\sum_{i=1}^4 E_{C,i}N_i^2,\qquad \sum_{i=1}^4N_i=0.
\end{equation}
$E_{C,i}$ denotes the charging energy of island $i$, and $N_i$ counts its excess Cooper pairs.
For Fig.~\ref{fig:spin-dependence}, the shared islands 1 and 2 have charging energies $1.05E_C$ and $0.95E_C$, respectively, while both outer islands have charging energy $E_C$.
Here $[\phi_i,N_j]=i\delta_{ij}$.

We diagonalize the Hamiltonian in a truncated integer-charge basis.
We compute the spin correlations and negativity from the reduced spin state $\rho_{AB}=\operatorname{Tr}_N\!\left(\lvert\psi_0\rangle\langle\psi_0\rvert\right)$, where the trace runs over the island charges.

\end{document}